# Candidate source of flux noise in SQUIDs: adsorbed oxygen molecules

Hui Wang,[1,2] Chuntai Shi,[2] Jun Hu,[2] Sungho Han,[2] Clare C. Yu[2] and R. Q. Wu[1,2]

[1]*State Key Laboratory of Surface Physics and Department of Physics, Fudan University, Shanghai 200433, CHINA*

[2]*Department of Physics and Astronomy, University of California, Irvine, CA 92697-4575, USA*

**Abstract:** A major obstacle to using SQUIDs as qubits is flux noise. We propose that the heretofore mysterious spins producing flux noise could be $O_2$ molecules adsorbed on the surface. Using density functional theory calculations, we find that an $O_2$ molecule adsorbed on an α-alumina surface has a magnetic moment of ~1.8 $\mu_B$. When the spin is oriented perpendicular to the axis of the O-O bond, the barrier to spin rotations is about 10 mK. Monte Carlo simulations of ferromagnetically coupled, anisotropic XY spins on a square lattice find 1/f magnetization noise, consistent with flux noise in Al SQUIDs.

Noise impairs the performance of a variety of devices based on superconducting circuits, e.g., photon detectors used in astrophysics [1], bolometers used in the search for dark matter [2], nanomechanical motion sensors [3], and quantum-limited parametric amplifiers [4]. Of particular interest are superconducting quantum interference devices (SQUIDs) [5] where low frequency 1/f magnetic flux noise [6] is one of the dominant sources of noise in superconducting qubits [7-10]. Experiments indicate that flux noise is produced by a high density (of order $5\times10^{17}$ m$^{-2}$) of fluctuating spins residing on the surface of normal metals [11] and superconductors [12, 13], though it is independent of the materials [6] Furthermore, experiments indicate that these spins are not independent, but rather may be clustered and have correlated fluctuations [14, 15].

A number of models of flux noise have been proposed [13, 16-19]. An early model of flux noise proposed that the spins are the magnetic moments of electrons in surface traps and that the spin orientation changes when an electron hops to a different trap [13]. Another model suggested that spin flips of paramagnetic dangling bonds occurred as a result of interactions with tunneling two-level systems [16]. Experimental indications of interactions between spins [12] led Faoro and Ioffe to suggest that flux noise is the result of spin diffusion via Ruderman-Kittel-Kasuya-Yosida (RKKY) interactions [17]. RKKY interactions between randomly placed spins produce spin glasses, and Monte Carlo simulations of Ising spin glass systems show that interacting spins produce 1/f flux and inductance noise in agreement with experiment [18].

The microscopic origin of these spins remains unclear. Choi *et al*. [20] proposed that they are electrons in localized states at the metal-insulator interface, though spins have also been found on the surface of the dielectric, aluminum oxide, without a metal present [11]. Density functional theory (DFT) calculations [21] on sapphire ($\alpha$-Al$_2$O$_3$), emulating the oxide layer that typically forms on surfaces of SQUIDs, indicate that thermodynamically stable charged vacancies are unlikely to be the source of flux noise because of the large energy differences associated with spin reorientation, though these energy differences decrease as the charge decreases. Lee *et al*. [21] used DFT to suggest that ambient molecules, such as OH, adsorbed on the surface could be the culprits, though the energy differences between different spin orientations is hundreds of degrees Kelvin, making thermal spin fluctuations unlikely.

Since SQUIDs are exposed to the atmosphere, we propose that the primary source of spins producing flux noise is O$_2$ molecules adsorbed on the surface. The free O$_2$ molecule has a spin triplet electronic configuration with a magnetic moment of 2.0 $\mu_B$ [22] and is strongly paramagnetic in its liquid phase. $O_2$ molecules absorbed on metal or oxide surfaces can form ordered lattices and exhibit exotic magnetic properties [23]. A natural question is whether they retain a large magnetic moment on the surface of metal oxides as well as on the surface of dielectric materials used to encapsulate SQUIDs [24]. If they do retain a large moment, it is important to know the associated magnetic anisotropy energies (MAEs) that are the energy barriers for spin reorientation and hence key to understanding thermal fluctuations. Due to the weak spin orbit coupling of oxygen, the MAEs of these systems are small, making them difficult to investigate theoretically and experimentally.

In this Letter, using systematic DFT calculations, we report that $O_2$ molecules with a surface density of $1.08 \times 10^{18}$ m$^{-2}$ have a large magnetic moment, 1.8 $\mu_B$/molecule, on an α-$Al_2O_3$ (0001) surface. These spin moments are weakly coupled and can reorient almost freely in a plane perpendicular to the O-O bond, with an energy barrier at the level of a few mK. Our Monte Carlo simulations on ferromagnetically coupled, anisotropic XY spins on a 2D square lattice suggest that they indeed produce 1/f magnetization noise, and hence an $O_2$ adlayer could be responsible for the flux noise found in SQUIDs. This would explain the long standing conundrum of why flux noise is independent of the materials [6].

Our DFT calculations were performed with the Vienna Ab-initio Simulation Package (VASP) [25-28], using the Perdew-Burke-Ernzerhof (PBE) functional [29] for the description of the exchange and correlation interactions among electrons. We treated O-2s2p and Al-3s3p shells as valence states and adopted the projector-augmented wave (PAW) pseudopotentials to represent the ionic cores [30, 31]. The energy cutoff of the plane-wave expansion was 500 eV. The spin-orbit coupling term was treated self-consistently using the non-collinear mode of VASP [32, 33], and the magnetic anisotropy energy was determined through either the torque or the total energy method [34, 35]. While our main results involved α-alumina, some test calculations were also carried out for γ-alumina thin films and ultra thin alumina on a NiAl (110) surface to investigate the effects of surface roughness and complex morphology [36]. To mimic sapphire $Al_2O_3$ (0001) surfaces, we constructed a slab model that consists of 18 atomic layers and a vacuum gap about 15 Å thick. In the lateral plane, we used a 2×2 supercell to dilute the adsorbates, corresponding to a surface density of $1.08 \times 10^{18}$ m$^{-2}$. The lattice constant in the lateral plane was fixed according to the optimized dimensions of bulk α-$Al_2O_3$ (a=b=4.81Å, c=13.12Å). A 11×11×1 Monkhorst-Pack [37] k-point mesh was used to sample the Brillouin zone. The criteria for structural optimization are (1) the atomic force on each atom is less than 0.01 eV/Å and (2) the energy convergence is better than $10^{-7}$ eV.

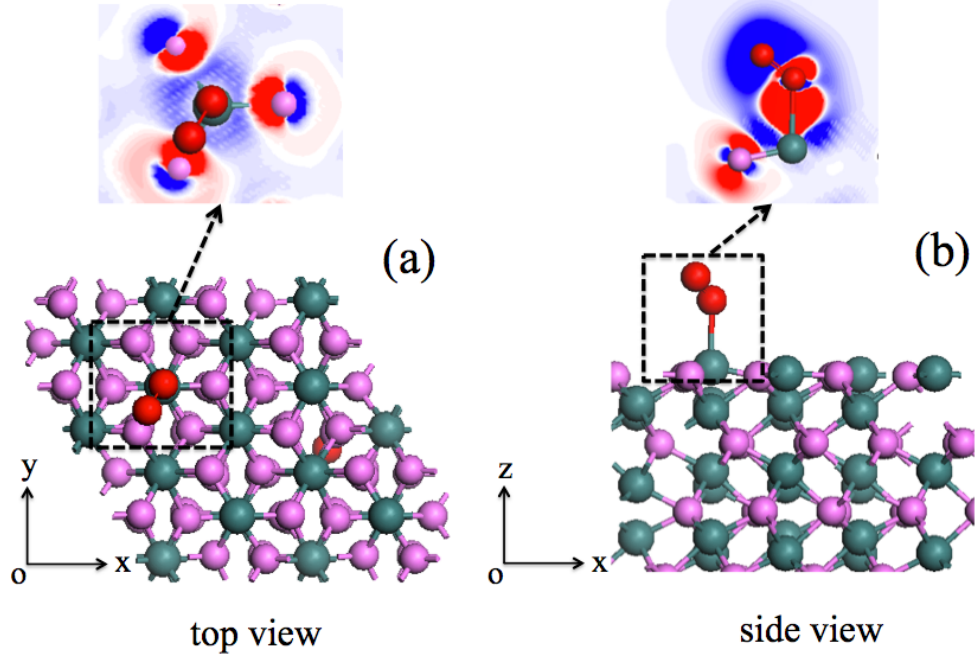

Fig. 1. (color online) Schematic atomic geometries of an $O_2$ molecule absorbed on an $Al_2O_3$ (0001) surface: (a) top view and (b) side view. The corresponding charge redistribution (the dash rectangular area) between an adsorbed $O_2$ molecule and the substrate is given in the insets: red and blue represent the charge accumulation and depletion at 0.005 $eV/Å^3$, respectively. Red balls and green balls represent the absorbed $O_2$ molecule and Al atoms, respectively. Magenta balls represent oxygen atoms in the $Al_2O_3$ lattice.

To describe the strength of $O_2$ adsorption, we define the binding energy per $O_2$ molecule as:

$$E_b = E_{O_2/Al_2O_3(0001)} - E_{Al_2O_3(0001)} - E_{O_2} \qquad (1)$$

where $E_{O_2/Al_2O_3(0001)}$ and $E_{Al_2O_3(0001)}$ are the total energies of the $Al_2O_3$ slab with and without the $O_2$ molecule on it. $E_{O_2}$ is the total energy of the free $O_2$ molecule in its gas phase. Through studies of various initial adsorption configurations, with the $O_2$ molecule being placed on top of O, Al, and O-O bridge sites, we found that the most preferential absorption site for the $O_2$ molecule is atop the Al site on the $Al_2O_3$ (0001) surface, with a binding energy of -0.15 eV. This indicates that the $O_2$-$Al_2O_3$ (0001) interaction is rather weak, which is understandable since the clean $Al_2O_3$ (0001) surface is known to be inert towards adsorbates [38, 39]. As shown by the red balls in Fig. 1(a) and Fig. 1(b), the absorbed $O_2$ molecule is tilted by about 55 degrees away from the surface normal. The optimized O-O bond length of 1.23 Å, which is close to the experimental value, 1.21 Å [40]. The nearest O-Al distance is 2.17 Å, and the Al atom underneath is pull up by about 0.34 Å from its position in the clean $\alpha$-$Al_2O_3$ (0001) surface, which is nevertheless still 0.50 Å lower than its bulk-like position [41].

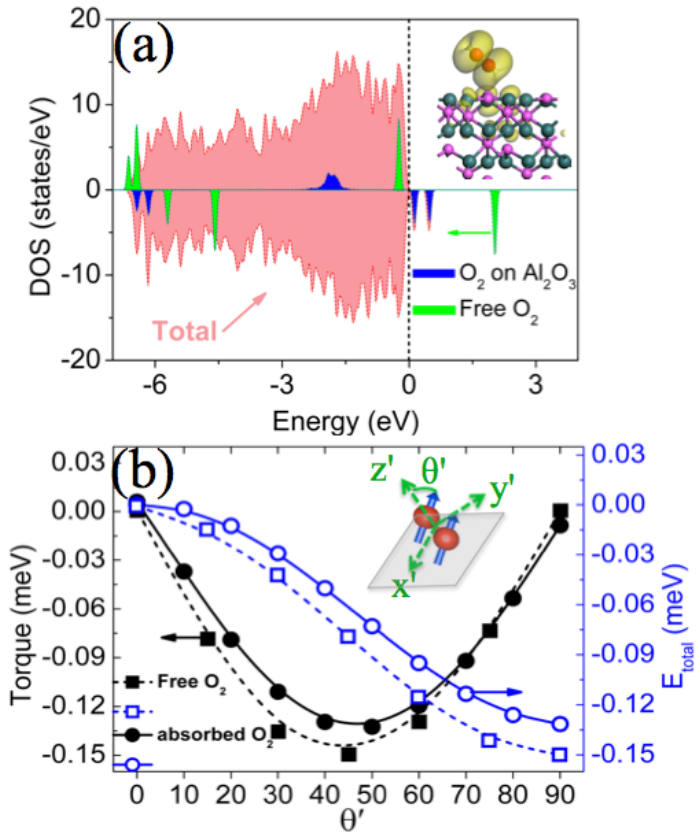

Fig. 2 (color online) (a) Projected density of states (PDOS) of the absorbed $O_2$ molecule on $Al_2O_3$ (0001), along with the total density of states of $O_2/Al_2O_3(0001)$ and the density of states of the free $O_2$ molecule. The positive and negative values correspond to states in the majority and minority spin channels, respectively. The inset gives the isosurfaces (olive) of the total spin density of $O_2/Al_2O_3(0001)$ at 0.005 $e/Å^3$. (b) Calculated torque and relative total energy ($E_{total}$) versus the spin orientation of the free (dashed lines) and adsorbed (solid lines) $O_2$ molecule. Inset defines the polar angle ($\theta'$) for the direction of the spin with respect to the $z'$ axis that lies along the O-O bond.

The total magnetic moment of each $O_2$ molecule on $Al_2O_3$ (0001) is 1.8 $\mu_B$, slightly smaller than that in its gas phase, 2.0 $\mu_B$. From the total density of states (TDOS) [peach background in Fig. 2(a)] and the projected density of states (PDOS) of the $O_2$ molecule [blue and green peaks in Fig. 2(a)], it is also evident that the pp$\pi$* orbitals of the $O_2$ molecule in the minority spin channel split into two separate peaks and shift down to the Fermi level from 2.0 eV for the free $O_2$ molecule. The small occupancy in the hybridized pp$\pi$* orbitals cause the charge rearrangement as depicted in the insets of Fig 1(a) and (b), and is responsible for the reduction of the magnetic moment of $O_2$. It appears that the lower oxygen atom in $O_2$ and the lattice oxygen atoms gain electrons from Al and the higher oxygen atom in $O_2$. The spin density of the absorbed $O_2$ molecule in the inset in Fig. 2(a) shows a donut feature of the pp$\pi$* orbital, similar to that of the free $O_2$ molecule. Meanwhile, the underlying Al and lattice O atoms are weakly magnetized, with small spin moments of 0.01 and 0.03 $\mu_B$, respectively.

The two key parameters for 1/f noise are the magnetic anisotropy energy (MAE) and the exchange interaction between $O_2$ molecules ($J_{ij}$). Our DFT calculations with 2×2 and 4×4 supercells indicate $O_2$ molecules order ferromagnetically on $Al_2O_3$(0001), with exchange energies of 0.14 meV (~1.6 K) for two oxygen molecules 4.8 Å apart, and 0.05 meV (~0.6 K) for a separation of 9.6 Å. It appears that the substrate plays a key role for the magnetic coupling between $O_2$ molecules, because calculations for free $O_2$ molecules gave smaller exchange values. As seen from the PDOS curves in Fig. S1 in the supplementary materials, the pp$\pi$* orbitals of antiferromagnetically (AFM) coupled $O_2$/$Al_2O_3$(0001) shift to higher energies compared to their counterparts in ferromagnetically (FM) coupled $O_2$/$Al_2O_3$(0001). This indicates a slightly smaller charge gain from Al when $O_2$ spins are AFM aligned compared to FM aligned. Meanwhile, the induced spin polarization on the lattice oxygen atoms between two $O_2$ molecules is also somewhat suppressed in the antiferromagnetic case. These factors facilitate the ferromagnetic coupling between $O_2$ on $Al_2O_3$(0001), as we will assume for the Monte Carlo simulations.

The determination of the small MAE of $O_2$ is still a challenge for DFT calculations. We calculated the torque τ (θ′) as a function of the polar angle θ′ of the spin moment with respect to the O-O bond (the z′ axis shown in the inset of Fig. 2(b): $\tau(\theta') = \frac{\partial E_{total}(\theta')}{\partial \theta'} = \sum_{occ} \langle \psi_{i,k} | \frac{\partial H_{SO}}{\partial \theta'} | \psi_{i,k} \rangle$ [34, 35], in steps of 15°, as illustrated in the inset. For the *free* $O_2$ molecule, $\tau$ follows the function -sin(2θ′) as shown by the dashed black line in Fig. 2(b). By integrating $\tau$ from 0 to θ′, we obtain the angle dependence of the total energy, $E_{total}$(θ′). Clearly, the lowest energy corresponds to the spin aligned perpendicular to the O-O bond (θ′=90$^0$), and the energy difference between θ′=0$^0$ and θ′=90$^0$ is 0.15 meV/$O_2$. Note that the spin rotation within the *x′y′*-plane has no energy barrier for the free $O_2$ molecule due to the cylindrical symmetry. Similarly, the torque associated with $O_2$/$Al_2O_3$ also follows -sin(2θ′) as shown by the solid black line in Fig. 2(b). The total energy decreases monotonically as the magnetic moment rotates away from the O-O bond towards the *x′y′* plane and the energy difference between θ′=0 and θ′=90º is 0.13 meV/$O_2$, slightly smaller than that of the free $O_2$ molecule. This MAE is sufficient to block the thermal spin fluctuations out of the *x′y′* plane toward the *z′* axis at temperatures below 1 K. Nevertheless, spin can rotate within the *x′y′* plane, and the corresponding energy barrier is the key to determining its contribution to magnetic noise. By using the torque and total energy methods, we found that this energy barrier is extremely small [about 1 μeV or 10 mK], almost at the limit of the precision that DFT can achieve for the determination of MAE, so rotation of spin within the *x′y′*-plane is unblocked.

Now that we know that the magnetic moments of $O_2$ molecules are weakly coupled on $Al_2O_3$(0001) and can easily rotate around the O-O bond, we want to see if they produce the 1/f noise observed in SQUIDs, rather than white noise (or a Lorentzian spectrum at low temperatures). So we performed Monte Carlo simulations of classical anisotropic XY spins. We will focus on exchange interactions between oxygen spins since dipolar and hyperfine [19] interaction energies much smaller, and describe the ferromagnetic nearest-neighbor exchange interactions with the Hamiltonian:

$$H = -\sum_{\langle i,j \rangle} J_{ij} \left( S_i^x S_j^x + S_i^y S_j^y \right) - A \sum_i \left( S_i^x \right)^2 \qquad (2)$$

where $S_i^x$ is the x-component of the spin on site *i and A* is the rescaled MAE. Without loss of generality, we choose the preferred anisotropic direction to be along the *x*-axis which we will refer to as the 'easy axis.' The length of the spins is 1. Since the SQUID surface is disordered and the oxygen molecules are adsorbed in random places, we chose ferromagnetic $J_{ij}$ > 0 from a Poisson-like distribution *P(J)* in the following way. First dimensionless integers $C_{ij}$ are drawn from a Poisson distribution with a mean of $<C_{ij}>$ = 5. Then $J_{ij} = 0.2J_0C_{ij}$, where the average coupling $<J_{ij}> = J_0 = 1$ sets the energy and temperature scale. For $A = 0$ and uniform coupling ($J_{ij} = 1$), we obtain the traditional 2D XY model which undergoes a Kosterlitz-Thouless phase transition [42] at $T_C \sim 1$ (the exact value of $T_C$ depends on the system size). We decided to use ferromagnetic couplings because (a) DFT finds ferromagnetic couplings; (b) the carrier density in the oxide is too low for spins to engage in RKKY interactions; and (c) there is experimental evidence for time reversal symmetry breaking consistent with surface spin ferromagnetism [14, 18]. We performed Monte Carlo simulations with the Metropolis algorithm on a 32x32 square lattice with periodic boundary conditions. In a trial move, a site and a trial angle between 0 and $2\pi$ are randomly chosen from a uniform distribution. At each temperature, the system is allowed to equilibrate for $10^6$ Monte Carlo steps per spin (MCS) before recording the time series for $M(t)$, the magnetization per spin, and for $E(t)$, the energy per spin. We then calculate the magnetization spectral density $S_M(\omega) = 2\int_{-\infty}^{\infty} dt e^{i\omega t} \langle \delta M(t) \delta M(0) \rangle$, where $\delta M(t) = [M(t) - \langle M \rangle]$. We normalize the noise power by setting the total noise power equal to $\sigma_M^2$, the variance of $M$: $S_{tot} = (1/N_\tau) \sum_{\omega=0}^{\omega_{max}} S_M(\omega) = \sigma_M^2$, where $N_\tau$ is the duration of the time series.

Our magnetization noise power results are shown in Fig. 3 (a) and (b). At high frequencies $S_M(f) \sim 1/f^\alpha$ where the noise exponent $\alpha$ varies from 0.3 to 2, depending on the temperature and anisotropy. At the Kosterlitz-Thouless transition (A=0 and T = $T_C \sim$ 1), our exponent is consistent with the expected value of [43] $\alpha = 1+(2-\eta)/z \sim 1.9$ with the critical exponent $\eta = ¼$ [44] and the dynamical critical exponent z $\sim$ 2 [45]. According to the actual values of J and A from the DFT calculations discussed above, the experimentally relevant parameters are T > 1.6 and A $\sim$ 0.01. In this regime, Fig. 3(c) and (d) show that the noise exponents range from 1.37 (for T=1.6) to 0.86 (for T=2.0) which is consistent with experimental values 0.58 to 1 [6, 9]. At low frequencies the noise is white due to a finite size effect [43]. We present additional results including the specific heat and susceptibility in the supplement.

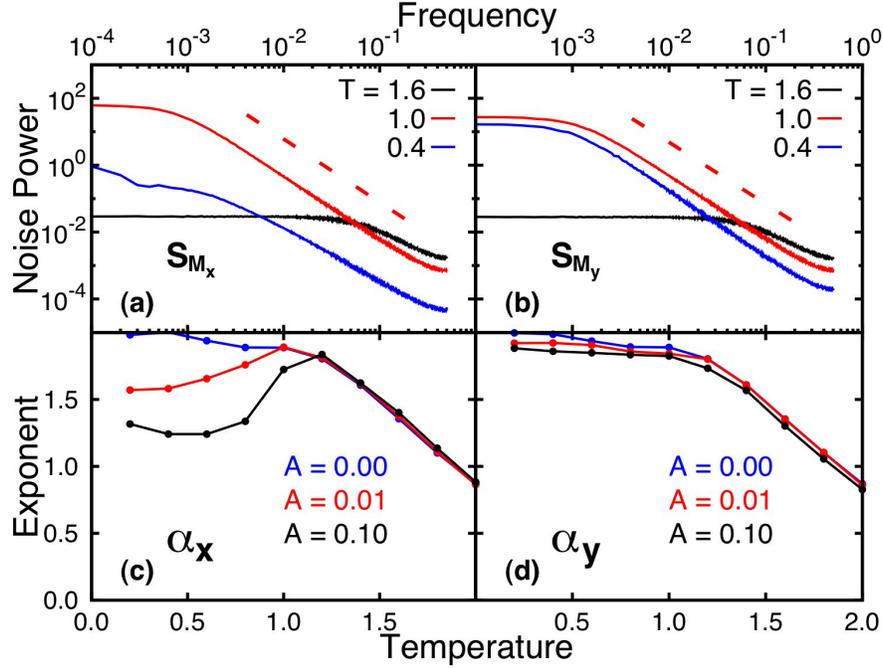

*Fig. 3 (color online): (a) and (b): Log-log plot of the magnetization noise power $S_M(f)$ vs. frequency (in units of 0.1/MCS) for (a) $M_x$ and (b) $M_y$ at various temperatures for A=0.01. The slopes of the dashed lines give the noise exponents α and are (a) 1.89 and (b) 1.84 for T=1. The noise spectra are taken from time series with $10^7$ MCS and averaged over 50 sample realizations of the couplings. (c) and (d): Noise exponents vs. T for various values of the anisotropy for (c) $M_x$ and (d) $M_y$.*

Now it is clear that the $O_2$ molecules on qubits are magnetic and can produce 1/f noise. While vacancies on the oxide surfaces may also produce local magnetic moments, they contribute much less to the noise since, as we show in the supplement, the formation energy for both Al and O vacancies are high (>2.4 eV) and hence their area density should be very low. Furthermore, none of vacancies induces a magnetic moment on the more complex γ-alumina surface (see supplement). Our recent calculations have also found that the x-ray magnetic circular dichroism (XMCD) spectrum of an $O_2$ adlayer has a sharp feature at the onset due to the transition from the 1s-shell to the characteristic 2π* orbitals of $O_2$, very different from that of O-vacancies. This offers a useful way for experimental verification in the future. The identification of $O_2$ adsorbates as the main source of magnetic noise has the important implication that one can reduce flux noise by protecting the surface with preoccupants such as $NH_3$, $N_2$ or CO. Our preliminary results indicate that the adsorption energy of $NH_3$ on sapphire is 1.8 eV per molecule, much higher than that of $O_2$, 0.15 eV.

In conclusion, systematic DFT calculations of $O_2/Al_2O_3$ (0001) demonstrate that the adsorbed $O_2$ molecule has a magnetic moment of ~1.8 $\mu_B$ and a small magnetic anisotropy energy of 10 mK. Monte Carlo simulations of ferromagnetically coupled anisotropic XY spins on a square lattice find 1/f magnetization noise, consistent with flux noise in Al SQUIDs. We thus propose that this could be the source of low frequency flux noise in SQUIDs. Unlike vacancies which may or may not produce magnetic moments, depending on the charge state and their local environment [21], adsorbed $O_2$ molecules have robust magnetic moments because of their weak interaction with the substrate. Furthermore, the experimentally estimated density of fluctuating spins, $5\times10^{17}$ m$^{-2}$ [11-13], is too high for vacancies, but is comparable to the surface density of $O_2$ adsorbates. Our results imply that flux noise could be substantially reduced by removing oxygen adsorbates from the surface of SQUIDs.

We thank John Clarke for helpful discussions. Work at UCI was supported by DOE-BES (Grant No. DE-FG02-05ER46237) and the Army Research Office (Grant No. W911NF-10-1-0494). Work at Fudan was supported by the Chinese National Science Foundation under grant 11474056 and the 1000 talent program of China. Computer simulations were performed with the U.S. Department of Energy Supercomputer Facility (NERSC).

# Supplementary Material on
# Candidate source of flux noise in SQUIDs: adsorbed oxygen molecules

Hui Wang,[1,2] Chuntai Shi,[2] Jun Hu,[2] Sungho Han,[2] Clare C. Yu[2] and R. Q. Wu[1,2]

[1]*State Key Laboratory of Surface Physics and Department of Physics, Fudan University, Shanghai 200433, CHINA*

[2]*Department of Physics and Astronomy, University of California, Irvine, CA 92697-4575, USA*

**DFT Results of the FM and AFM $O_2/Al_2O_3(0001)$:** To understand the mechanism of magnetic exchange coupling between $O_2$ molecules on $Al_2O_3(0001)$, we plot their PDOS curves and isosurfaces of the spin density. The main feature in the PDOS is the shift of $2\pi^*$ states to the higher energy side in the AFM case compared to the FM case. This indicates a smaller amount of electrons on $O_2$ in the AFM $O_2/Al_2O_3(0001)$ or, in the other word, slightly less charge gain of $O_2$ from the Al atom underneath because of the change of magnetic ordering. The isosurfaces of the spin density show that the induced spin polarization around the lattice oxygen atoms between $O_2$ molecules is also suppressed in the AFM case.

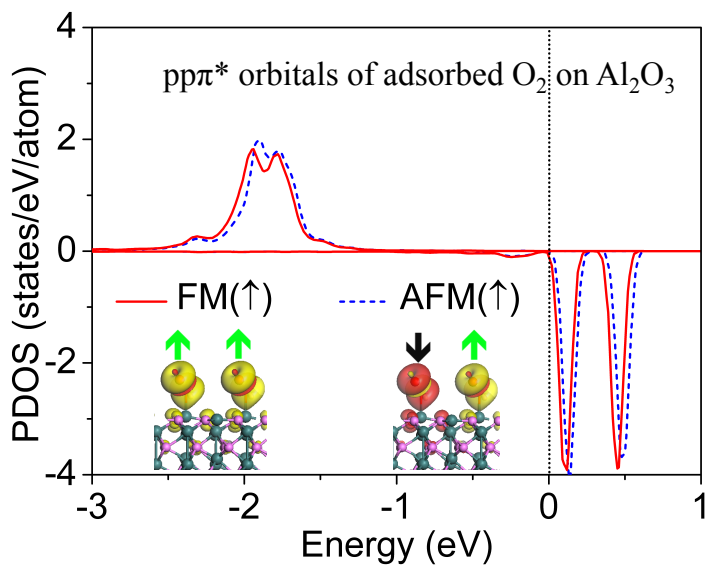

*Figure S1. Projected density of states (PDOS) of the ppπ\* orbitals of an absorbed O$_2$ molecule on Al$_2$O$_3$ (0001). The positive and negative values correspond to the majority and minority spin states, respectively. The insets give the isosurfaces of spin density of the FM and AFM O$_2$/Al$_2$O$_3$(0001) at 0.005 e/Å$^3$ (olive for positive spin density and red for negative spin density).*

**Results of DFT simulations on Vacancies:** Another possible set of candidates for the spins producing flux noise are the magnetic moments of native defects in aluminum oxide surfaces. So, in addition to adsorbed O$_2$ molecules, we used DFT to investigate various vacancies on an α-alumina surface as spin candidates. Various vacancies (e.g., O vacancies, Al vacancies, Al-O di-vacancies and Al-O-O tri-vacancies) in an α-alumina surface can also be magnetic, as has been found for "nonmagnetic" oxides such as MgO [1], Al$_2$O$_3$ [2], and HfO$_2$ [3]. As we describe below, we find that neutral Al vacancies and Al-O divacancies produce spin polarization on the surface of sapphire, with small magnetic anisotropy energies. However, these moments are sensitive to their local environment and are not present in γ-alumina which is closer in structure to native aluminum oxide. Thus we propose that the spins associated with adsorbed O$_2$ are more likely to be responsible for the flux noise found in SQUIDs.

To study vacancies, we performed electronic structure calculations as described in the main text with a 2×2 supercell to dilute the density of adsorbates and vacancies, corresponding to a surface density of $1.08 \times 10^{18}$ m$^{-2}$. The formation energies $E_f$ of vacancies are defined by

$$E_f = E(V) - E(Al_2O_3) + \sum n_V \mu_V \qquad (2)$$

where E(V) and E(Al$_2$O$_3$) are the total energies of an Al$_2$O$_3$ supercell with and without vacancies, respectively. $\mu_V$ and $n_V$ represent, respectively, the chemical potential and the number of aluminum and oxygen atoms that are removed. Under equilibrium growth conditions, the chemical potentials of Al ($\mu_{Al}$) and O ($\mu_O$) should obey the constraint: $2\mu_{Al} + 3\mu_O = \mu_{Al_2O_3}$, where $\mu_{Al_2O_3}$ is the chemical potential per Al$_2$O$_3$ unit in bulk sapphire.

We considered several types of vacancies in the 2×2 supercell of α-Al$_2$O$_3$ (0001), including a single O vacancy (V$_O$), a single Al vacancy (V$_{Al}$), an Al-O di-vacancy (V$_{Al-O}$) and an Al-O-O tri-vacancy (V$_{Al-O-O}$), to see whether they can be magnetic. From the large formation energies listed in Table 1, we can exclude the presence of a single O vacancy and an Al-O-O tri-vacancy. Furthermore, our DFT calculations suggest that the magnetic moment induced by a single O vacancy is negligible (<0.1 μ$_B$).

Al vacancies and Al-O di-vacancies may occur in appreciable concentrations in $Al_2O_3$ samples due to their relatively lower formation energies, $E_f(V_{Al})$ = 2.8 eV and $E_f(V_{Al-O})$ = 2.4 eV, as shown in Table 1. The calculated magnetic moment induced by a single Al vacancy is rather large, about 3.0 $\mu_B$ (with about 0.82 $\mu_B$ from each O atom adjacent to the vacancy) as listed in Table 1, consistent with previous studies [4]. The TDOS in Fig. S2 shows sharp peaks in the minority spin channel at about 1 eV above Fermi level. After the removal of an Al atom, one pp$\sigma$ orbital from each adjacent O atom becomes unoccupied in the minority spin channel, which is the reason why the spin density in the inset resides in O-Al pp$\sigma$ orbitals. For Al-O di-vacancies, the induced magnetic moment is about 1.0 $\mu_B$, consisting primarily of about 0.45 $\mu_B$ residing in the dangling bonds on each neighboring O site. Thus the origin of the magnetization for the Al-O di-vacancy is not much different from that of a single Al vacancy.

*TABLE 1. The formation energy ($E_f$), the total magnetic moment (M) and the magnetic anisotropy energy (MAE) with respect to different axes for a single O vacancy, an Al-O-O tri-vacancy, a single Al vacancy and an Al-O di-vacancy.*

| Vacancies | $E_f$ (eV) | M ($\mu_B$) | MAEs ($\mu eV$) | |
| --- | --- | --- | --- | --- |
|  |  |  | $E_x - E_z$ | $E_y - E_z$ |
| $V_O$ | 5.65 | 0.04 | -- | -- |
| $V_{Al-O-O}$ | 5.45 | 1.0 | -- | -- |
| $V_{Al}$ | 2.80 | 3.0 | -0.6 | -0.6 |
| $V_{Al-O}$ | 2.40 | 1.0 | -24.0 | 3.0 |

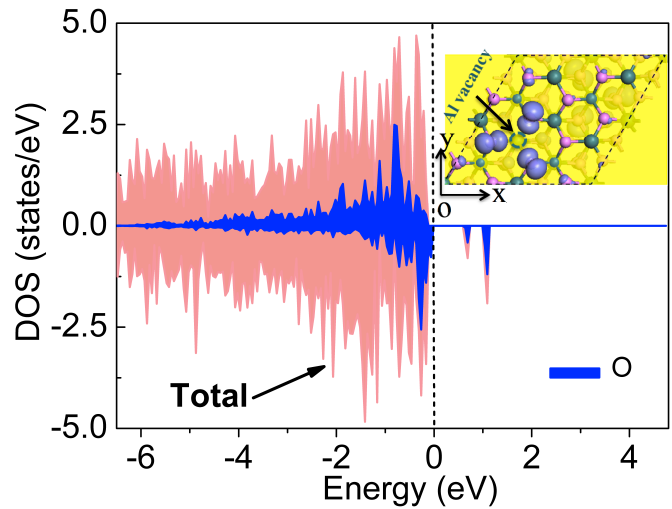

*Figure S2: Total (peach-shaded area) and projected (blue area) density of states of the $Al_2O_3$ surface with a single Al vacancy. Positive and negative values indicate states in the majority and minority spin channels, respectively. The inset shows the isosurfaces (purple spheres) of its spin density at 0.05 e/Å³. Green circle indicates the position of the Al vacancy.*

To see if vacancy-induced magnetization on α-$Al_2O_3$ (0001) contributes to the flux noise on metal oxide surfaces, we calculated the MAEs of a single Al vacancy and an Al-O di-vacancy. For the single Al vacancy, the energy barrier for spin flipping between the three axes is less than 0.6 µeV (see Table 1), indicating that the spin can reorient easily. Although the spin orientation of Al-O di-vacancies is relatively more stable, with an MAE of 24 µeV (see Table 1), spin fluctuations can still occur at very low temperatures on α-$Al_2O_3$ (0001). Therefore, these spins can also cause flux noise if they exist. However, sapphire may not be as representative of ambient aluminum oxide as γ-alumina. So we also investigated the magnetization of Al vacancies and Al-O di-vacancies in γ-alumina and ultra thin alumina films. Interestingly, none of these vacancies induces a magnetic moment on these more complex surfaces. The magnetic moments from native defects on alumina appear to be very sensitive to the charge state and the local environment, which is consistent with the work of Lee *et al.* [4] on the effect of charge on the magnetization of vacancies in α-$Al_2O_3$ (0001). This makes it unlikely that vacancies are the source of flux noise. In contrast, the magnetization of $O_2$ molecules adsorbed on well-separated Al cations is much more robust. Furthermore, the experimentally estimated density of fluctuating spins, $5 \times 10^{17}$ m$^{-2}$ [5, 6], is too high for vacancies, but is comparable to the surface density of $O_2$ adsorbates.

In conclusion, systematic DFT calculations of $O_2$/$Al_2O_3$ (0001) demonstrate that, unlike vacancies which may or may not produce magnetic moments, depending on the charge state and their local environment [4], adsorbed $O_2$ molecules have robust magnetic moments because of their weak interaction with the substrate.

**Results of Monte Carlo Simulations:** Here we present further results of our Monte Carlo simulations of ferromagnetically coupled, anisotropic XY spins on a square lattice.

To calculate the specific heat $C_V$, we record the time series of the energy E per particle and then use the formula $C_V = N\sigma_E^2/(k_BT^2)$ where N is the number of spins and $\sigma_E^2$ is the variance of the energy per particle. Figure S3 shows the specific heat as a function of temperature. The peaks indicate the estimated transition temperature in these finite size systems.

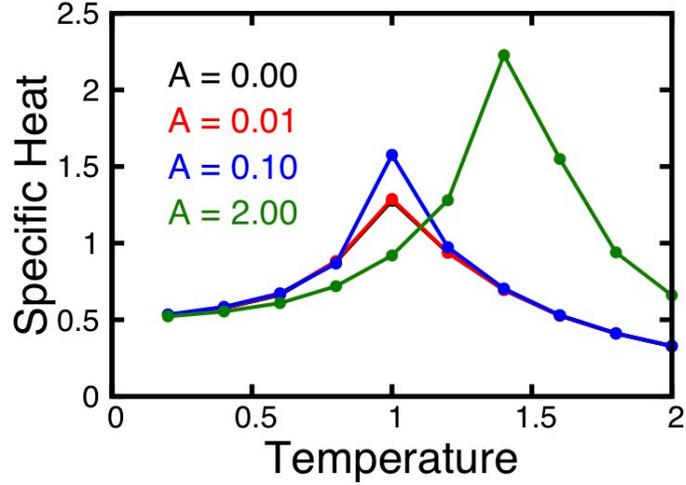

*Figure S3: Specific heat versus temperature for various values of the anisotropy energy A. The curves for A=0 and A=0.01 are superposed. System size is a 32 x 32 square lattice. Results were averaged over 50 sample realizations of the couplings and simulations were run for $10^7$ MCS.*

To calculate the susceptibility, we record the time series of the magnetization M per spin for the x and y components of the magnetization. Then we use the formula $\chi_{aa} = N\sigma_M^2/(k_B T)$ where $a$ is the x or y component and $\sigma_M^2 = \left(\langle M_a^2 \rangle - \langle M_a \rangle^2\right)$ is the variance of the appropriate component of the magnetization per spin. Figure S4 shows the x- and y-components of the susceptibility as a function of temperature.

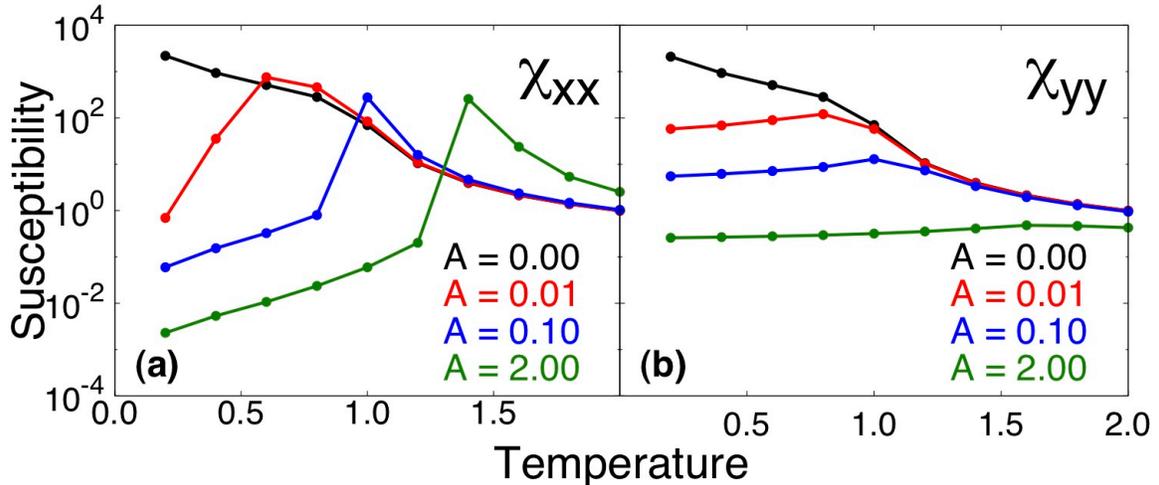

*Figure S4: Susceptibilities $\chi_{xx}$ and $\chi_{yy}$ versus temperature for various anisotropy energies. System size is a 32 x 32 square lattice. Results were averaged over 50 sample realizations of the couplings and simulations were run for $10^7$ MCS.*

Note that this susceptibility differs from the often-used susceptibility [7] $\chi_{|M|} = \frac{N}{k_B T}\left(\langle M^2 \rangle - \langle |M| \rangle^2\right)$ which is associated with |M|, the magnitude of the magnetization per spin. $\chi_{aa}$ and $\chi_{|M|}$ differ for a variety of reasons. First of all, $\chi_{|M|}$ cannot be written as the derivative of the magnetization with respect to a physical field. Second, the variance depends on the square of the mean, and the mean is zero at all temperatures for the individual components of the magnetization but nonzero below the transition for the magnitude of the magnetization. A peak in $\chi_{|M|}$ identifies the transition temperature $T_C$ [7]. At temperatures above $T_C$, a good estimate for the physical susceptibility is $\chi_1 = \frac{N}{k_B T}\langle M^2 \rangle$ since <M> = 0 in the thermodynamic limit [7]. Notice that if <$M_x$> = <$M_y$> = 0, then $\chi_1 = \chi_{xx} + \chi_{yy}$. We show a comparison of $\chi_{xx} + \chi_{yy}$, $\chi_1$ and $\chi_{|M|}$ versus temperature for A = 0.1 in Figure S5.

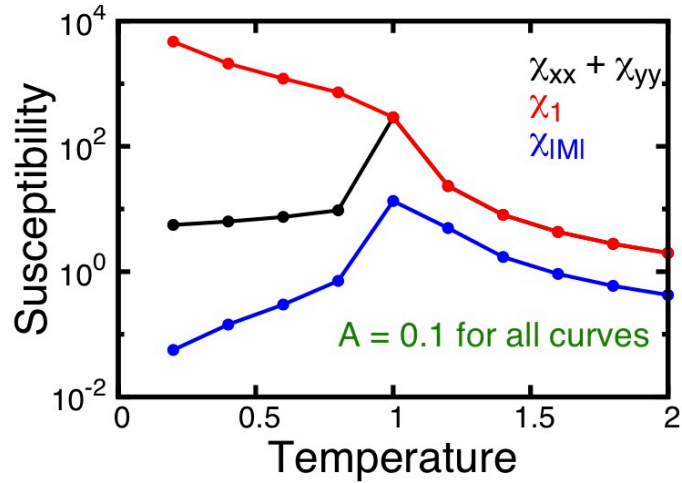

*Figure S5: The susceptibilities $\chi_{xx} + \chi_{yy}$, $\chi_1$ and $\chi_{|M|}$ versus temperature for A = 0.1.*

In Figure S6 we show $\chi_{|M|}$ as a function of temperature for various anisotropy energies. The peaks give an estimate of the transition temperatures in these finite size systems.

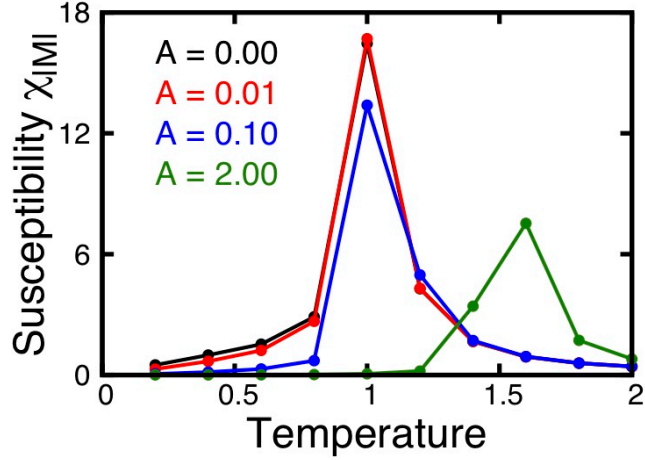

*Figure S6: The susceptibility $\chi_{|M|}$ versus temperature for various anisotropy energies.*

The noise spectra as a function of temperature for various values of the anisotropy are shown in Figure S7 (a) and (b). The noise spectra are normalized so that the total noise power integrated over frequency is equal to the variance, i.e., $S_{tot} = (1/N_\tau) \sum_{\omega=0}^{\omega_{max}} S_M(\omega) = \sigma_M^2$ where $N_\tau$ is the duration of the time series used to calculate $S_M(\omega)$.

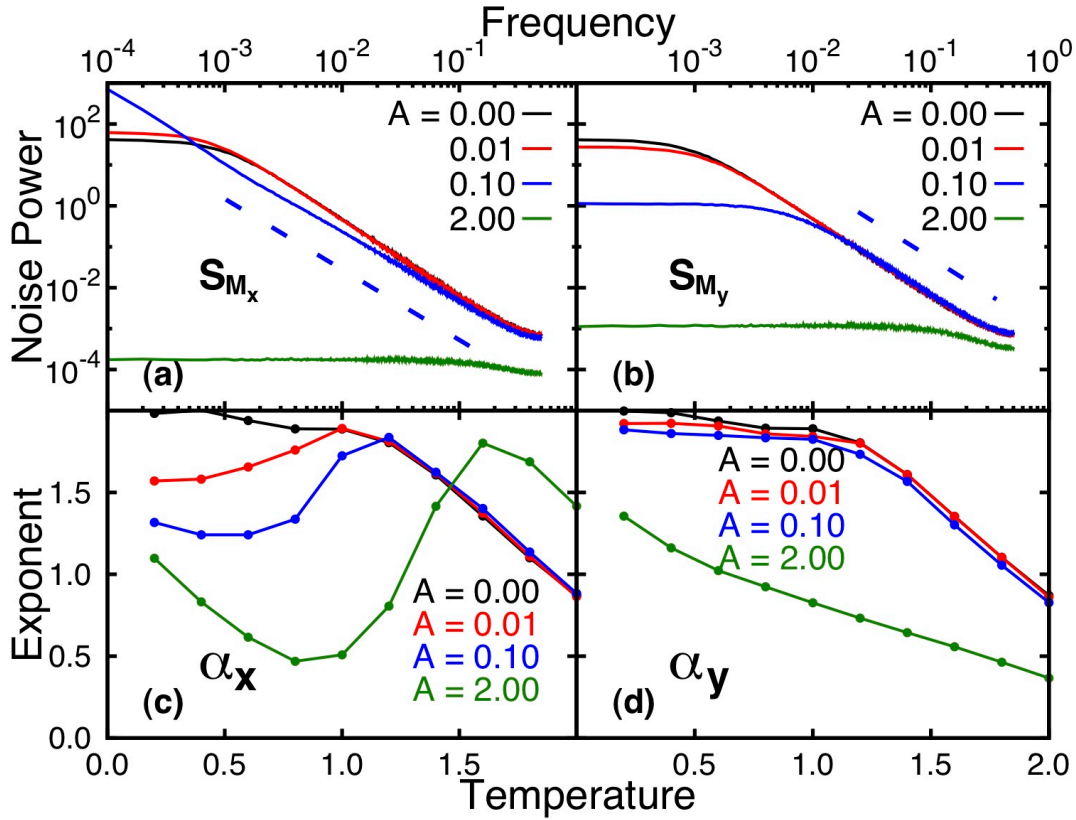

*Figure S7: (a) and (b): Log-log plot of the magnetization noise power $S_M(f)$ vs. frequency (in units of 0.1/MCS) for (a) $M_x$ and (b) $M_y$ at anisotropies at $T = 1$. The slopes of the blue dashed lines give the noise exponents α and are (a) 1.72 and (b) 1.82 for $A = 0.1$. The noise spectra are taken from time series with $10^7$ MCS and averaged over 50 sample realizations of the couplings. (c) and (d): Noise exponents vs. T for various values of the anisotropy for (c) $M_x$ and (d) $M_y$.*